\documentclass[twocolumn,prb,showpacs]{revtex4}

\usepackage{graphicx}
\usepackage{rotating}
\usepackage{amsmath}
\usepackage{amsfonts}
\usepackage{amssymb}
\usepackage{enumerate}
\usepackage{longtable}
\usepackage{dcolumn}
\usepackage{bm}
\usepackage{amsmath}
\usepackage{placeins}

\begin{document}

\newcommand{\mto}{${\rm MgTi_2O_4}$}
\newcommand{\mts}{${\rm MgTi_2O_4}~$}
\newcommand{\nn}{\nonumber}
\newcommand{\MTO}{MgTi$_2$O$_4$}
\newcommand{\dxy}{\ensuremath{d_{xy}}}
\newcommand{\dxz}{\ensuremath{d_{zx}}}
\newcommand{\dyz}{\ensuremath{d_{yz}}}
\newcommand{\tg}{\ensuremath{t_{2g}}}
\newcommand{\eg}{\ensuremath{e_{g}}}
\newcommand{\ag}{\ensuremath{a_{1g}}}
\newcommand{\egp}{\ensuremath{e'_{g}}}

\title{Orbital-spin order and the origin of structural distortion in
${\rm \bf MgTi_2O_4}$}

\author{S. Leoni$^1$}
\author{A.N. Yaresko$^2$}
\author{N. Perkins$^3$}
\author{H. Rosner$^1$}
\author{L. Craco$^1$}

\affiliation{$^1$Max-Planck-Institut f\"ur Chemische Physik fester
Stoffe, D-01187 Dresden, Germany}
\affiliation{$^2$ Max-Planck-Institut f\"ur Festk\"orperforschung, Heisenbergstra\ss e 1, D-70569 Stuttgart, Germany}
\affiliation{$^3$ University of Wisconsin - Madison, 1150 University Avenue Madison, WI 53706-1390, USA}

\date{\rm\today}

\begin{abstract}
We analyze electronic, magnetic, and structural properties of the spinel
compound \mts using the local density approximation+U method. We show how
\mts undergoes to a canted orbital-spin ordered state, where charge,
spin and
orbital degrees of freedom are frozen in a geometrically frustrated network
by electron interactions. In our picture orbital order stabilize the
magnetic ground state and controls the degree of structural distortions.
The latter is dynamically derived from the cubic structure in the correlated LDA+U potential.
Our ground-state theory provides a consistent picture for the dimerized phase
of \mto, and might be applicable to frustrated materials in general.
\end{abstract}


\maketitle

\section{Introduction}

Over the last years, a lot of progress has been achieved in the
study of  transition metal oxides on frustrated lattices. Interest
in these systems stems from the richness of their novel properties:
the unexpected variety of ordered states and transitions between
them, and the complexity of the underlying physics. Transition metal
oxide (TMO) spinels AB$_2$O$_4$,  with magnetic B ions forming a 
pyrochlore lattice, give the unique possibility to
explore how the natural tendency of correlated systems to develop
magnetic, orbital, and charge order is effected by geometrical
frustration.~\cite{varios,Verwey} The best studied example
of TMO spinels, and historically the first one, is  magnetite ${\rm
Fe_3O_4}$ which shows a high Curie temperature and undergoes the
Verwey transition at $T_V\simeq 120$~K.~\cite{Verwey}

\begin{figure}[t]
\includegraphics[width=\columnwidth]{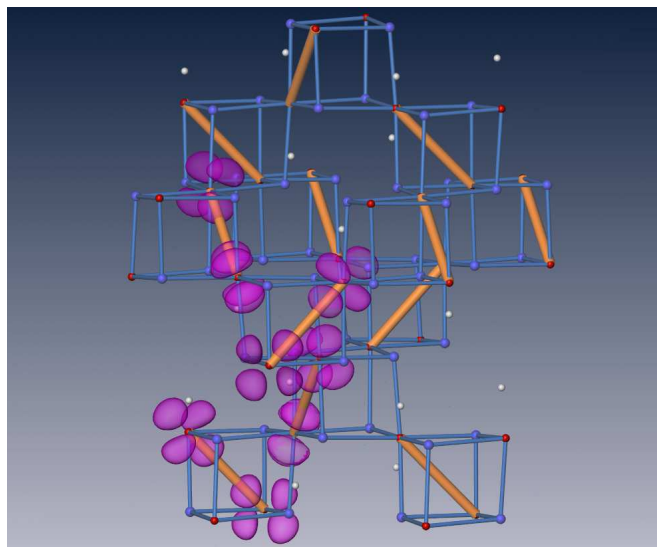}
\caption{
Orbital order in the tetragonal crystal structure of \mts within the
LDA+U framework ($U=3$~eV). The electron density of the most populated 
$t_{2g}$ orbital is plotted. Lattice constants are taken from Ref.~\cite{radaelli}. 
The orange bonds represent the shortest
bonds in the distorted structure, and the corresponding atoms are displayed
in red (Ti), blue (O) and white (Mg).}
\label{fig1}
\end{figure}

Recently, B-spinel \mto, which is characterized by a pyrochlore lattice of
Ti$^{3+}$ magnetic ions 
with one single electron in the $t_{2g}$-manifold, has attracted much
attention due to a very peculiar 
phase transition from a metallic to a spin-singlet insulating phase near

$T_C \approx $~260~K.~\cite{isobe} The signature of the insulating state
is the optical gap ($\approx$ 0.25~eV at T=10~K) observed in optical
conductivity spectra.~\cite{Zhou} Taken together,
dc-resistivity~\cite{isobe,Good} and optical~\cite{Zhou,Popo} measurements
consistently indicate that \mts undergoes a sharp metal-insulator
transition (MIT) with no sign of Drude weight at low frequencies below $T_C$.
The MIT is accompanied by a structural transition
from cubic to tetragonal symmetry, with a concomitant
drop of the magnetic susceptibility and a resistivity jump below $T_C$.
Neutron diffraction and X-ray measurements~\cite{radaelli} indicate
spin-dimerization: Ti-Ti dimers are formed in a helical pattern in the
spin-singlet state. Further structural refinements~\cite{RadNJP}
reveal that Ti-ions move away from the center of the TiO$_6$ octahedron. 
In the low-T phase two out of six Ti-Ti bonds gets closer, suggesting the formation
of chemical dimers. These findings have suggested a removal of the
pyrochlore degeneracy by a 1D helical dimerization of the spin pattern,
with spin-singlets (dimers) located at short bonds.

There were several theoretical attempts to understand the nature of
spin-dimerization and the origin of the MIT
in \mto. Khomskii and Mizokawa~\cite{Khoms}  assumed that the system
is close to an itinerant state, and explained the formation of the
nonmagnetic spin-singlet state by exploiting the concept of a {\it
orbitally driven Peierls state}, leading to the formation of quasi
1D bands. In this case, the magnetic changes across the
structural transition in \mts can be understood within the picture
of 1D Peierls transition driven by the ordering (Fig.~\ref{fig1}) of the
$d_{xz}$ and $d_{yz}$ orbitals. This picture is consistent with
B3LYP functional (GGA) calculations~\cite{radaelli} for the tetragonal
phase of \mto, showing that the $xz$ and $yz$ orbitals are occupied
while the $xy$ states are pushed to high energies.

Another approach  is based on the assumption that the
low-temperature tetragonal phase of \mts is  Mott
insulator.~\cite{natalia} In this case, the ground state  of \mts
can be found by studying an effective low-energy spin-orbital
super-exchange Hamiltonian. It was shown in~\cite{natalia} that the
orbital degrees of freedom in \mts modulates the spin exchange
couplings, providing an explanation for the helical spin-singlet
pattern observed in.~\cite{radaelli} It appears that the minimum
energy configuration corresponds to such an orbital ordering for
which  the maximum number of spin-singlet dimers is formed.
However, there are various dimer coverings of the pyrochlore lattice
which have the same energy. As  in spin-Peierls systems, the
increase of magnetic energy  gain can be achieved by the shortening
of bonds, where dimers are situated. Therefore, each type of dimer
coverings corresponding to a particular  orbital ordering  induces
a different distortion of the lattice, which costs a different elastic
energy. The ground state is chosen simply by finding a  minimum
energy state, which in case of \mts corresponds  to the minimal
enlargement of the unit cell.

However, no theoretical studies have been yet performed in order to
understand the origin of the MIT and the
formation of the spin-singlet state. 
Is it the intrinsic lower
dimensionality of \mts that causes the formation of the  spin-singlet
state, or is the driving mechanism of different type?

In this work, in order to address these questions, we perform
quantitative investigation of the band structure of \mto. Using
local (spin) density approximation plus Hubbard U, L(S)DA+U,
approach,~\cite{LDA+U} we show how an explicit incorporation of
electronic correlations allows for a realistic description of the
insulating ground state of \mto. Our results reveal that electronic
correlations are fundamental in stabilizing the dimerized ground
state of \mto. We show that MIT is driven by
correlation (via LDA+U) induced orbital order (OO) rearrangement.
Therein, U controls band splitting towards an orbital-insulating state {\it without}
full orbital polarization (OP).
In our picture OO stabilizes the spin-singlet ground state,
which in turn controls the degree of structural distortions. This
finding is consistent with the superexchange spin-orbital
description.~\cite{natalia} The opening of the electronic band gap
is understood in terms of an orbital-selective MIT~\cite{Anisi} on a
quasi 1D network.

\section{Results and Discussion}

To elucidate the interplay between spin, orbital, and charge
degrees of freedom in \mts we perform {\it correlated}
scalar-relativistic band-structure calculations using local (spin)
density approximation plus Hubbard U, L(S)DA+U,
approach.~\cite{LDA+U} We employ the linear muffin-tin orbitals
(LMTO) scheme in the atomic sphere approximation, with combined
correction terms.~\cite{ok} 

\begin{figure}[t]
\includegraphics[width=3.0in]{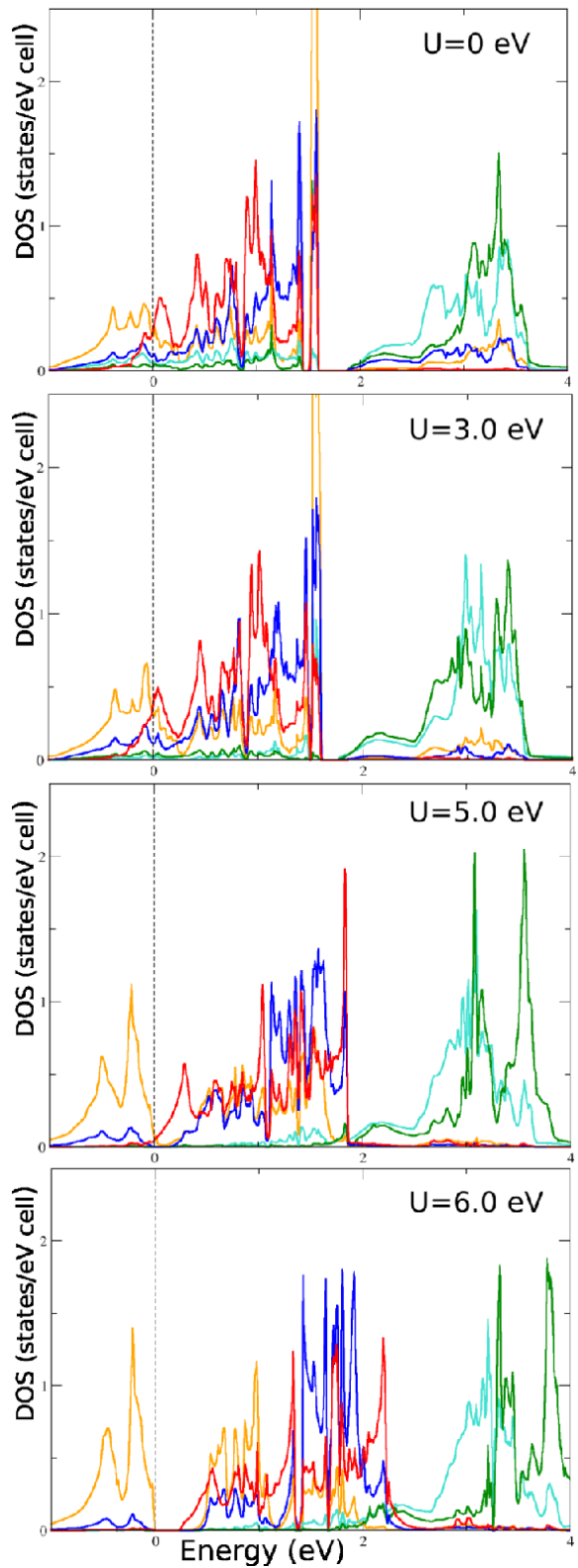}
\caption{
LDA+U orbital resolved DOS for the tetragonal 
$(P4_12_12)$~\cite{radaelli} phase of \mto: $U=0$~eV (top left), $U=3$~eV
(top right), $U=5$~eV (botton left), and $U=6$~eV (botton right).
Notice the orbital selective nature of the correlation induced
metal-insulator transition between 5~eV $<U<$ 6~eV solutions.}
\label{fig2}
\end{figure}
\FloatBarrier

Self-consistency is reached by
performing calculations on a 12x12x10 {\bf k}-mesh for the Brillouin zone
integration. 
Experimental atomic positions \cite{radaelli} for the tetragonal $(P4_12_12)$
low-$T$ phase were used in the calculations.
The radii of the atomic spheres were chosen as
$r=2.4789$ (Mg), $r=2.0022, 2.0106$ (O1, O2), and $r=2.5857$ (Ti) a.u. in order to minimize
their overlap.

Our results for the paramagnetic (LDA+$U$) phase of \mto~are presented in Fig.~\ref{fig2}. Therein 
the evolution of the $3d^1$ correlated density of states
(DOS) is shown. In cubic spinel oxides the $d$-electron orbital sector splits
into the low-energy $t_{2g}$ and high-energy $e_{g}$ orbitals. Below
we will denote the $t_{2g}$ orbitals  in the tetragonal phase as
$XY$, $XZ$, and $YZ$, which differ from the $\dxy$, $\dyz$ and
$\dxz$ in the cubic phase (see details in \cite{orbitals}). The
corresponding DOS are plotted in red ($XY$), orange ($XZ$), and
blue ($YZ$) in Fig.~\ref{fig2}. The DOS with dominant $e_g$ contribution
are shown by green and cyan curves. Due to partial oxygen
relaxation, the crystal field has also a small trigonal component,
which further splits the $t_{2g}$ manifold  into $a_{1g}$ and
$e'_{g}$ sectors .~\cite{symorb}  We note that the $XZ$ orbitals
directed along one of the short Ti--Ti bonds can only be formed as a
linear combination of the $a_{1g}$ and $e'_{g}$ orbitals. According
to LDA results for cubic \mts the occupation of the $e'_{g}$ states
is appreciably higher than that of the $a_{1g}$ one.
Already in LDA calculations for the dimerized phase, the degeneracy of the
$e'_{g}$ orbitals is lifted as a consequence of the local distortion of
TiO$_6$ octahedra and overall tetragonal symmetry. One of them (orange curve
in Fig.\ 2) acquires $XZ$ character and becomes more populated than the $YZ$
one (blue). The $XY$ orbital (red), originating from the $a_{1g}$ state,
forms a peak just above the Fermi level ($E_F$), its occupation being less
than that of the other two.
This picture does not change much in the weak correlation regime ($U\lesssim$3
eV).

With increasing $U$ ($3 ~{\rm eV} \le U \le 5$ ~eV) transfer of spectral weight
significantly modifies the weakly correlated scenario such that the $XZ$ orbital
is now almost half filled ($n_{XZ}>0.8$), whereas $YZ$ and $XY$ states are
pushed towards the $e_g$ bands, reducing the $t_{2g}-e_g$ charge
transfer gap.
In this large $U$ regime the overall electronic bandwidth is enhanced, 
accompanied by a
reduction of the DOS near $E_F$ and selection of a single orbital channel to form the insulating state.
We observe a clear tendency towards a pseudo-gap formation in the
$XZ$ and $YZ$ channels ($U = 5$ ~eV), while the $XY$ orbitals are pushed to
higher energies. 
This indicates OP as the precursor of the MIT.
By further increasing $U$ we obtain an insulating state at $U=6$~eV. Our paramagnetic
insulating solution is characterized by an almost half filled $XZ$
band, with appreciable splitting between occupied bonding and unoccupied
antibonding states.

The bonding-antibonding splitting observed at large $U \ge 5$~eV
(Fig.~\ref{fig2}) aids the formation of the charge gap. The insulating
state can be viewed as a consequence of strong hopping between the $XZ$
orbitals along the short Ti--Ti bonds, leading to
robust singlet character of these bonds (see discussion below) along the
$c$-axis. 
Thus, according to our results \mts undergoes an orbital
selective MIT,~\cite{Anisi} which is caused by correlation assisted
orbital rehybridization.~\cite{kot_rehy}

\begin{figure}[tb!]
\includegraphics[width=3.1in]{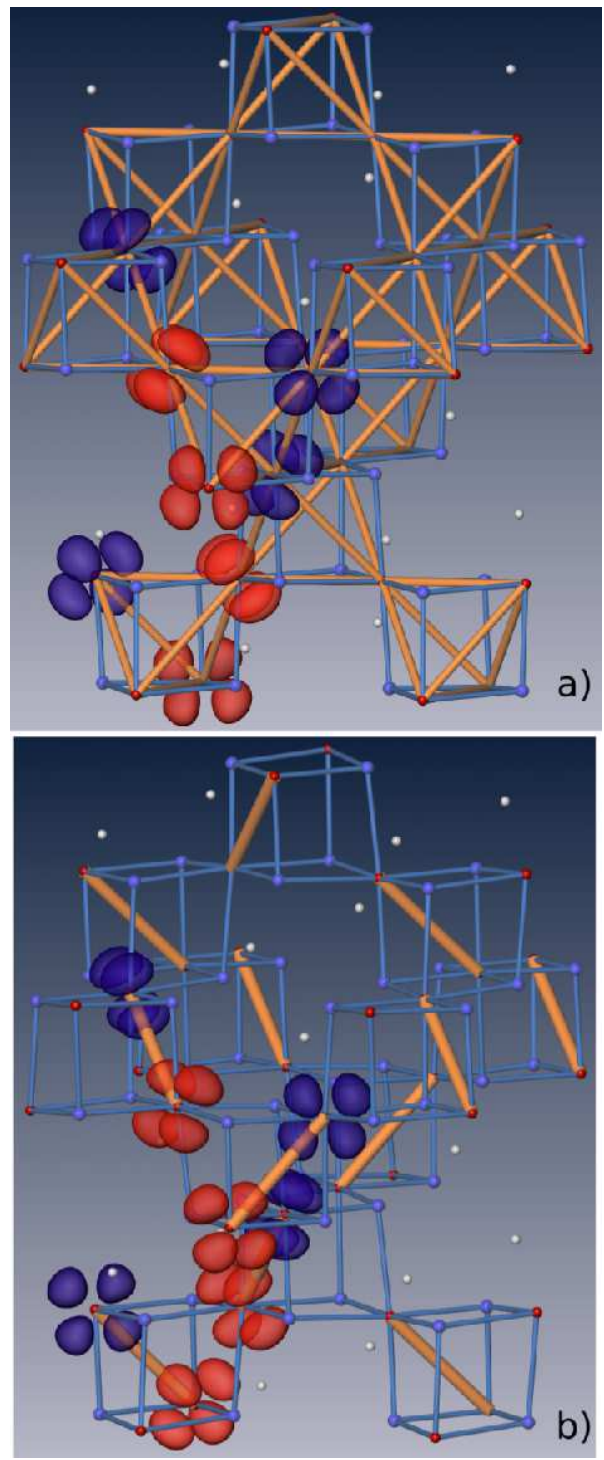}
\caption{
Orbital ordering of \mts as obtained within LSDA+U ($U$=3~eV) in the
antiferro-pseudocubic (a, upper panel) and antiferro-dimerized
phase (b, lower panel). Red and blue colors denote the ground-state orbital with
different $(\uparrow,\downarrow)$ spins.}
\label{fig3}
\end{figure}
\FloatBarrier

The changes in OO and crystal structure are shown in
Fig.~\ref{fig1}. Consistent with~\cite{Khoms} ferro-orbital order (FOO) along
the short Ti--Ti bonds, which stabilizes the molecular bond (orange)
formation, starts to develop already for $U \le 3$~eV. The spiralling of the
dimer bonds along the $c$-axis is visible in Fig.~\ref{fig1}. For
$U=3$~eV we obtain FOO, with the ground state orbital
pointing through bond's direction.  

\begin{figure}[tb!]
\includegraphics[width=\columnwidth]{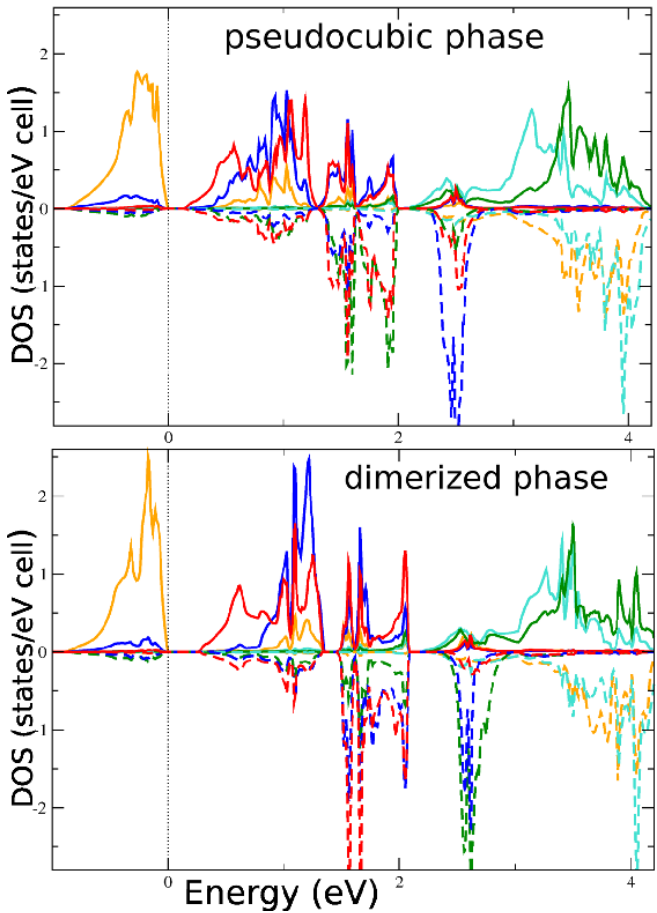}
\caption{
LSDA+U orbital resolved DOS for $U=3$~eV in the antiferro-pseudocubic
phase (upper panel) and in the antiferro-dimerized phase (lower panel) of \mto.
Within dimers, DOS of different sites are mirror images with respect to spin space.}

\label{fig4}
\end{figure}

We now turn our attention to orbital, spin and charge
responses in the magnetically ordered state. In Figs.~\ref{fig3}
and~\ref{fig4} we display our LSDA+U ($U=3$~eV) results for an idealized
{\it pseudocubic} (upper panel) pyrochlore structure and the low-T
tetragonal structure (lower panel).~\cite{radaelli}

In the idealized pseudocubic (PC) phase all atoms are arranged as in 
the high-T cubic structure supplemented by a reduced $(P4_12_12)$
tetragonal space group. This combination of atomic arrangements and
crystal symmetries allows for the discovering of novel magnetic and
orbital reorientations in the undistorted pyrochlore structure of \mto,
which are in principle allowed in the vicinity of
the MIT point.~\cite{natalia}

\FloatBarrier

The orbital resolved DOS of the PC structure (Fig.~\ref{fig4}) shows large OP 
and a small charge gap between the ground $(XZ)$ and the first
excited $XY$ orbital. Notice the dramatic rearrangement of the minority
states of the ground state orbital, which are shifted to energies above
3~eV. 

Within the tetragonal metric of the low-T phase we find a similar evolution
for the magnetically-ordered electronic DOS as for the pseudocubic
regime. 

In the lower panel of Fig.~\ref{fig3} we display the
ground-state OO in the distorted phase of \mto. 
The LSDA+U solution gives 
the same antiferromagnetic arrangement of spins
along the short Ti--Ti bonds 
and in the $a,b$ plane of the pyrochlore
lattice, as found for the PC structure. 
The lower panel of Fig.\ 3 shows, however, that the canting of the $XZ$ orbitals
away from the Ti--Ti bond plane becomes signicantly smaller when the low-$T$
structural distortions are taken into account. Also, the average Ti--O
distance in the $XZ$ plane (2.068 \AA) is appreciably larger than in the
$YZ$ (2.052 \AA) and $XY$ (2.041 \AA) planes, which means that the occupation
of the $XZ$ state becomes energetically more favorable.
This suggests a novel scenario for quasi
1D chains~\cite{Khoms} in the strongly frustrated network of \mto, where the
crystal-structure itself gets modified by the onset of OO. The small changes
in the atomic positions of the Ti-ions shown in Fig.~\ref{fig3} additionally
suggest that crystal structure transformations are coupled to
the correlation-induced orbital re-orientation. 

\begin{figure}[t]
\includegraphics[width=3.0in]{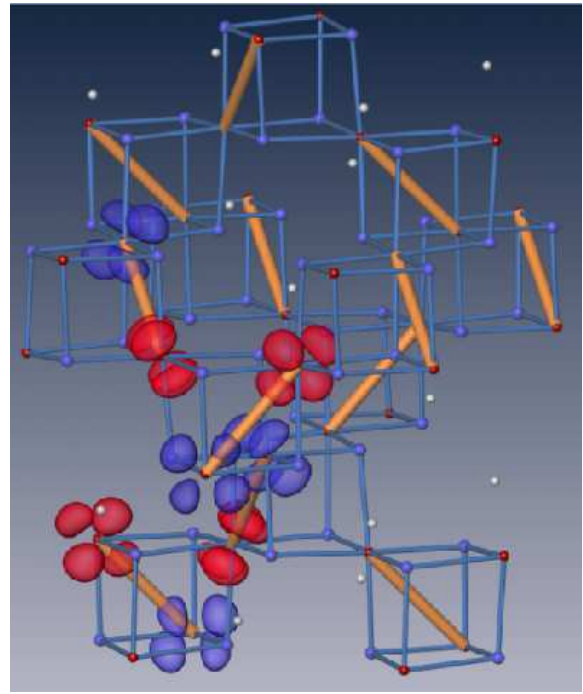}
\caption{
Additional magnetically ordered LDA+U solution ($U=3$~eV) with the same FOO as the ground-state solution, however with FM spin order in the $a,b$ plane and AFM spin order along $c$.  Red and blue colors denote the orbitals with
different $(\uparrow,\downarrow)$ spins.}
\label{fig6}
\end{figure}
\FloatBarrier

\begin{figure}[tb!]
\includegraphics[width=3.1in]{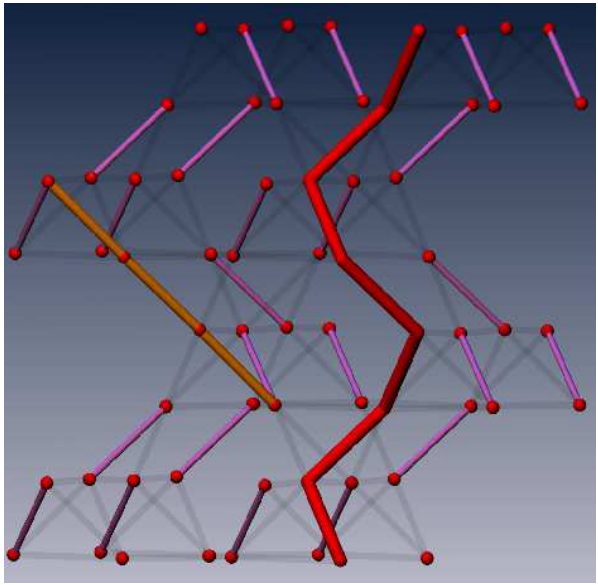}
\caption{
($U=3$~eV). Ti-Ti shortest dimers are represented in mauve.
Tetramers and spirals ( only one representative is shown in orange and red, respectively)
arises from distorting the pyrochlore lattice (black transparent bonds) from cubic to tetragonal.
Only Ti atoms are displayed for clarity.}
\label{fig5}
\end{figure}
\FloatBarrier

Another magnetically ordered LSDA+U solution (Fig.~\ref{fig6}) with the same FOO but FM spin order in
the $a,b$ plane and AFM one along the $c$ direction can also be obtained for both the PC and dimerized structures.
The order along the short Ti--Ti bonds remains FOO with AFM spin order,  
however, the Ti spins at the ends of intermediate Ti-Ti bonds with
AFOO ($XZ$--$YZ$) in the $c$ direction are antiparallel. As a consequence, 
the total energy of this solution is higher than the
ground state solution shown in Fig.\ \ref{fig3}, in which the order along intermediate
bonds is AFOO with FM spin order.

The dimerization of AFM spin chains and
the formation of  spin-singlets located on
short bonds explains the crystal structure  of \mts, which can be seen as a
collection of helices running along the $c$
axis. This helix is  formed by an alternation of short-long-short-long
bonds as found in.~\cite{radaelli} As a consequence of the selective
bond shortening superstructure features like  tetramers can be recognized (Fig.~\ref{fig5}).

The static calculations have provided evidence on the role of electronic correlation
in the formation of the spin-singlet state. To further unravel the coupling between structural rearrangement and correlation-assisted OO, we have performed full structure relaxation in the correlated LDA+U potential. Starting from the cubic structure with pyrochlore sublattice, Parrinello-Rahman \cite{PR} (PR) structure relaxation \cite{espresso} was performed. The Lagrangian formulation of PR allows for constant-pressure relaxation simulations, as variation of size and shape of the simulation box are allowed. This is typically used in connection with pressure-induced phase transitions. Here $U$ is acting as some sort of ``pressure'' on the orbital channels, whose rehybridization brings structural changes about. To properly capture this effect it is thus important to dispose of a molecular dynamics setup allowing for unbiased geometry changes. For $U=3$~eV structural lock-in into the tetragonal dimerized phase is achieved with selective shortening of a subset of bonds ({\it molecular dimers}) as displayed in Fig.~\ref{fig5}. The role of local Coulomb interaction $U$ as symmetry-reducing agent through selection of one orbital channel is apparent in the frame of the dynamic calculations.
The found dimensional reduction selects thus a particular topological order among the various "degenerated valence bond configurations", allowed in the quantum dimer framework of frustrated lattices,\cite{Senthil} in agreement with experiments.

\section{Conclusion}

In conclusion, we have studied the ground state orbital, charge and
magnetic properties of \mts using the LDA+U technique. Our results
suggests an orbital selective~\cite{Anisi} picture for the
metal-insulator transition of \mto, which is driven by large
spectral weight transfer due to correlations (via LDA+U). This in
turn introduces a new type of bonding-antibonding
splitting near the Fermi energy, which is characterized by large
orbital polarization of excited orbitals. Using the LSDA+U results
we have computed the orbital order pattern in both high-T
pseudocubic and dimerized phase. The insulating low-T phase is
shown to be driven by electron correlations in concert with spin and
orbital order. The existence of short Ti-Ti distances along the $c$-axis
follows as a consequence of spin dimerization on bonds (molecular
dimers). Finally, consistent with x-ray diffraction
experiments~\cite{radaelli} the resulting superstructure features
(tetramers, spirals) are shown to be carried by correlation
over orbital order, explaining magnetic and structural properties of
\mts in its low-T insulating phase.

\acknowledgements

S.L. thanks the Swiss National Foundation for financial support.
L.C. and H.R. acknowledge support from the Emmy Noether-Programm of the DFG.
Computational time provided by ZIH, Dresden is aknowledged.


\begin{references}
\bibitem{symorb} In the cubic frame $a_{1g}$ and $e'_g$ states can be written
  as: $d_{a_{1g}} = (\dxy+\dyz+\dxz)/\sqrt{3}$, $d_{1e'_{g}} =
  (2\dxy-\dyz-\dxz)/\sqrt{6}$, $d_{2e'_{g}} = (\dyz-\dxz)/\sqrt{2}$.
\end{references}
\end{document}